\documentclass[english,twocolumn]{revtex4}
\usepackage[latin1]{inputenc}
\usepackage{amsmath}
\usepackage{graphicx}
\usepackage{amssymb}
\usepackage{dcolumn}
\makeatletter
\usepackage{dcolumn}

\makeatletter

\usepackage{dcolumn}

\usepackage{bm}

\usepackage{mathrsfs}

\makeatother

\usepackage{babel}
\makeatother
\begin{document}

 \title{A combined R-matrix eigenstate basis set and finite-differences propagation
method for the time-dependent Schr\"{od}dinger equation: the one-electron
case}

\author{L. A. A. Nikolopoulos, J.S. Parker and K. T. Taylor}
\affiliation{Centre for Theoretical Atomic and Optical Physics, Department of
Applied Mathematics and Theoretical Physics, The Queen's University
of Belfast, BT7 1NN, Belfast, UK}

\date{\today}
\begin{abstract}
In this work we present the theoretical framework for the solution
of the time-dependent Schr\"{o}dinger equation (TDSE) 
of atomic and molecular systems
under strong electromagnetic fields with the configuration space of
the electron's coordinates separated over two regions,
that is regions $I$ and $II$. In region $I$ the solution of
the TDSE is obtained by an R-matrix basis set representation of the
time-dependent wavefunction. In region $II$ a grid representation
of the wavefunction is considered and propagation in space and time
is obtained through the finite-differences method. 
It appears this is the first time a combination
of basis set and grid methods has been put forward for tackling
multi-region time-dependent problems. In both regions,
a high-order explicit scheme is employed for the time propagation.
While, in a purely hydrogenic system no approximation is involved
due to this separation, in multi-electron systems the validity and
the usefulness of the present method relies on the basic assumption
of R-matrix theory, namely that beyond a certain distance (encompassing
region $I$) a single ejected electron is distinguishable from the other
electrons of the multi-electron system and evolves there (region II) effectively as a one-electron system.
The method is developed in detail for single active electron systems and applied
to the exemplar case of the hydrogen atom in an intense laser field. 
\end{abstract}
\maketitle

\section{Introduction}
\label{sec:1}
Exploration of the fundamental processes that occur when atomic and molecular systems are subject to extreme
conditions is currently a major research area. Experimentally, such
processes are realized by strong and/or short intense laser pulses
radiating at infrared wavelengths \cite{posthumus:2004,zeidler:2006}  
and have recently been utilised at a more practical level for reconstruction of nuclear probability
distributions, visualisation of molecular orbitals, alignment of molecules
as well as production of high-order harmonics which in turn are used
for the generation of ultra short fields at the attosecond scale \cite{Niikura02,Niikura03,alnaser:183202,itatani:2004,alnaser:163002,PhysRevLett.89.013001,zeidler:2006}.

Theoretically, it is a huge task to treat the exact time-dependent
(TD) response of a multi-electron system subject to a strong electromagnetic
(EM) field by \textit{ab initio} methods. In response to extensive experimental 
achievements using high-intensity Ti:Sapphire laser sources
in the long wavelength regime, many theoretical studies employed the
strong-field approximation  where the influence of the Coulomb
potential on the ejected electron wave function is neglected in favour
of the external field. A more sophisticated approach that adopts the
single-active-electron (SAE) approximation was also applied to the
atomic case~\cite{kulander:1987b}. SAE models where one reduces
the dimensionality of the multi-electron problem by freezing the most
tightly bound electrons have proven to be very useful in cases where
multiple electronic excitations are insignificant, and the SAE approximation
is probably the most widely used approach when studying phenomena
such as single ionization, above-threshold ionization (ATI) and high-harmonic
generation (HHG).

For systems of only two electrons, such as the negative hydrogen ion, helium,
molecular hydrogen, direct, ab-initio, solutions of the time-dependent Schr\"{o}dinger equation (TDSE) have 
appeared in the early nineties (for a review see ref. \cite{lambropoulos:1998}). 
Since then, the computational power has increased steadily and as a result these methods have reached
a high level of accuracy, efficiency and reliability, tackling successfully
the very demanding theoretical problem, of single and double ionization
of helium at 390 and/or 780 nm \cite{parker:2000,parker:2006}.

Recently, the construction of FEL sources which deliver brilliant
radiation in the soft- and (in the immediate future) hard X-ray regime  have initiated new challenges in the field of  
atomic and molecular physics \cite{moller:2005,sorokin:2006}. However, in contrast to what occurs with conventional laser 
sources, more than a single electron at a time responds to short wavelength FEL light 
and  X-ray FEL light will interact preferentially with the inner-most electrons, 
residing closer to the system's core, rather than with the valence ones.
An immediate consequence of the above property is that theories such
as the SAE and models not taking into account interelectronic interactions
at a sufficient level are inadequate to describe the processes involved.
Moreover, high-order harmonic generation (HOHG) techniques are nowadays
able to create pulses of subfemtosecond duration. Given that
relaxation processes, such as Auger transitions, of the bound electrons
are of the order of a femtosecond or less it can be concluded that the
short time-variation of the EM field requires approaches where multi-electron
dynamics can be reliably described. 

Given our intention to study multi-electron systems under intense EM
ultrashort fields, there is considerable importance in the development
of computationally tractable methods able to treat multi-electron systems
with the least approximations possible. Such approaches have been developed
in atomic and molecular physics studies, and include variants of
time-dependent Hartree-Fock (TDHF) ~\cite{kulander:1987b}. 
Though a vast number of theoretical efforts in the spirit of TDHF
\cite{kulander:1988,pindzola:1991,kulander:1991,pindzola:1993,pindzola:1995}
have appeared, even some extensions to include correlation between
the electrons, the question of how much and under what conditions
correlation beyond the Hartree-Fock model is important still remains
unanswered. The underlying reason is the difficulties introduced by
the nonlinear nature of the TDHF equations in combination with the
fact that the single-configuration ansatz and the excitation process
induced by the EM field are inconsistent. Improvements of the restricted
Hartree-Fock ansatz and inclusion of exchange effects appear to be
possible solutions to overcome such problems, although the applications
so far are only in one-dimensional (1D) models 
\cite{horbatsch:1992,dahlen:2001,bandrauk:2006,hu:1998,scrinzi:2005}.

An alternative \textit{ab-initio} approach capable of treating multi-electron
systems is R-matrix theory, with the basic formulation appeared first
in the context of nuclear theory, and later on applied in the field
of atomic physics (\cite{burke:1971,burke:1975a,taylor:1975}). Traditionally,
R-matrix theory is a theory where time is not involved in the study
of the collision or photoionization processes. Variants of R-matrix
theories and computational codes have been applied to an impressive
number of systems, over the last 40 years \cite{burke:1993}. With
the advent of strong and/or short laser pulse technology an early
application of R-matrix theory to multiphoton processes appeared in
the form of a Floquet expansion of the driven time-dependent wavefunction
\cite{burke:1991}. Although able to treat the field non-perturbatively,
the R-matrix Floquet approach cannot be considered as a fully TDSE
solution methodology since it is only suited to laser pulses containing
many cycles.

Similarly, the appearance of high power sources at the short wavelength
regime has led a number of theoretical groups to develop TDSE approaches
based on R-matrix theory (\cite{hart:2007,guan:2007,hart:2008}), 
with the first work to this end appearing some years ago \cite{burke:1997}. 
The basic assumption of R-matrix theory is very well suited to the physical
situation of the photoionization process involved in light-matter
interaction. Under strong radiation any system will ionize
either multiply or singly. In the regime of single ionization
the ejected electron, after some time, depending on its distance from
the core, can be safely identified as distinguishable from the other
electrons. In R-matrix theory this is taken into account through
the division of configuration space into two regions where, in the
inner region (region $I$), all interelectronic and exchange effects
between all the electrons are treated, while in the outer region (region $II$)
the ejected electron evolves effectively as a one-electron system
under the influence of the residual core and the potential due to the remaining electrons.
Thus in the outer region no matter what particular process has taken
place the system wavefunction consists entirely of that of the wavefunction
of the ejected electron.

The purpose of this work is two-fold. The first is to pursue development of a method
which meets the above requirements for more complex systems than one-
and two- electron systems and where atomic structure plays an important
role in the processes. For this, a method based on R-matrix basis
eigenstates appears to be tractable due to its success in describing
such complex systems. Second, and equally important is the issue of
efficiency and accuracy. It is inevitable that the demands of the
calculations will make the study of such problems computationally
very demanding. 
Finite-differences with high-order explicit time propagators \cite{smyth:1998} although difficult to use throughout
configuration space in a direct extension to multi-electron systems,
have proven to be very efficient and accurate in solving
the TDSE for one- and two-electron systems.  In fact the HELIUM code \cite{smyth:1998} using such methods to
solve the TDSE fully for a two-electron atom exposed to intense laser fields is able to run 
with high efficiency in both computation and communication over many thousands of cores on
the largest supercomputers presently available.  This established efficiency makes their
implementation for the outer region in our present approach a very reasonable one. 
In region $I$, an
R-matrix basis set is used to propagate the multi-electron wavefunction
while in region $II$ amounting effectively to a one-electron
problem, a finite-difference high-order propagation algorithm is used.
Since to the best of our knowledge no such attempt has appeared, namely
the propagation of the TDSE in a combined basis and grid representation
of the TD wavefunction, we consider it essential to set out carefully in detail
the basics of the method, free from complications arising from multi-electron
considerations. Thus, we provide below the details of the
method and its usefulness for one-electron systems and present results
for the hydrogen system where accurate ab-initio methods, to compare
with, are available to us. 

The paper is organized as follows. In Sec. ~\ref{sec:2} we give an overview of the basic 
ideas and principles. Section~\ref{sec:345} is the key section of this paper and there we set 
out in detail the theoretical formulation for a 
one electron system. 
In Sec. ~\ref{sec:6} we apply the method to the 
hydrogen atom in an intense laser field which serves as an exemplar. We have relegated to appendices 
some of the more technical details. Finally we set out some conclusions and perspectives with regard to
the new method in Sec. ~\ref{sec:7}.  Atomic units are 
used ($m=\hbar=|e|=a_{0}=1$) throughout.

\section{Overview of the basic ideas and principles}
\label{sec:2} 
As mentioned briefly in the introduction, the basic
assumption of R-matrix theory for the outer-region wavefunction
allows the derivation of a TDSE (in the outer region), where only
one electron is involved reducing the dimensionality of the problem
there to its minimum, namely to at most three, thus simplifying the computational
problem considerably. To put this in a more quantitative fashion,
let us recall the $(N+1)-$ electron wavefunction beyond a certain distance,
say $b$ (taken as the inner boundary of the outer region $II$) \cite{burke:1993}: 
\begin{equation}
\psi({\bf \tilde{r}}_{N},{\bf r};t)=\sum_{\gamma}\Phi_{\gamma}
({\bf \tilde{r}}_{N};\hat{r},\sigma_{N+1})\frac{1}{r}f_{\gamma}(r,t)\qquad r\ge b,
\label{eq:outer-rmx-wf}
\end{equation}
 with ${\bf \tilde{r}}_{N}=({\bf r}_{1},{\bf r}_{2},..,{\bf r}_{N})$, $r_i \le b, i=1,2,..,N$
and ${\bf r}={\bf r}_{N+1}$. The $\Phi_{\gamma}({\bf \tilde{r}}_{N};\hat{r},\sigma_{N+1})$
are channel functions formed by coupling the target states of the
residual atomic system $\phi_{\gamma}({\bf \tilde{r}}_{N})$, described
by the Hamiltonian $H_{N}({\bf \tilde{r}}_{N})$ and the angular and
spin quantum numbers of the ejected electron. The radial motion of
the ejected electron (in the $\gamma$-channel) is described by the
radial channel functions $f_{\gamma}(r,t)$. The absence of the antisymmetrization
operator is essential in the above expansion since it relies on the
ejected electron and the remaining $N-$electrons occupying different
portions of configuration space, thus making the ejected electron
distinguishable from the others. Let us now consider the TDSE of the
above system, in an external time-dependent radiation field.
By writing the Hamiltonian for the field-free $(N+1)-$ electron system as $H({\bf \tilde{r}_{N},{\bf r}})=-\nabla_{r}^{2}/2+H_{N}({\bf \tilde{r}}_{N})+V({\bf \tilde{r}_{N}},{\bf r})$
we end up with the following form for the TDSE:
 \begin{equation}
i\frac{\partial}{\partial t}\psi({\bf \tilde{r}}_{N},{\bf r};t)
=
\left[H({\bf \tilde{r}}_{N},{\bf r})+D({\bf \tilde{r}_{N}},{\bf r},t)\right]\psi({\bf \tilde{r}}_{N},{\bf r};t),
\label{eq:tdse}
\end{equation}
with $D({\bf \tilde{r}}_{N},{\bf r},t)$ denoting the interaction operator
between the system and the external field, in the dipole approximation.
Projection of the known channel states $\Phi_{\gamma}$ onto the TDSE
and integration over ${\bf \tilde{r}}_{N}$ and $\hat{r},\sigma_{N+1}$
results in the following set of coupled partial differential equation
for the radial motion in channels $\gamma$, 
\begin{equation}
i\frac{\partial}{\partial t}f_{\gamma}(r,t) =
\hat{h}_{\gamma}(r)f_{\gamma}(r,t)+\sum_{\gamma^{\prime}}\hat{D}_{\gamma,\gamma^{\prime}}(r,t)f_{\gamma^{\prime}}(r,t).
\label{eq:tdse_radial_1}
\end{equation}
 By properly ordering the radial channel functions $f_{\gamma}(t)$
into a column vector ${\bf F}(t)$ and the evolution operators $\hat{h}_{\gamma}$
and $\hat{D}_{\gamma,\gamma^{\prime}}$ into a square matrix $\hat{{\bf H}}(r,t)$
we may (in the outer region $II$) rewrite the TDSE of the ejected electron
of \textit{any multi-electron system} in the case of single ionization
as,
 \begin{equation}
i\frac{d{\bf F}}{dt}(r,t)=\hat{{\bf H}}(r,t){\bf F}(r,t)\qquad r\ge b,
\end{equation}
 this equation having essentially the form of the one-electron TDSE. It is exactly
this last equation, no matter how the inner region is treated, that
allows us to utilize any propagation technique in the outer region
$II$ of configuration space, which may have already
been applied to one-electron ionization.

On the other hand, in the inner region an eigenstate representation
of the TD wavefunction will result in a TDSE where only two dynamical
quantities are needed to be provided for the forward propagation in
time of the solution, namely eigenenergies and transition matrix
elements between the system's eigenstates. The key point in this case
is that the whole information about the exact nature of the system
described in the inner region, whether multi-electron or not, is
contained in the values of the energies and the transition matrix elements
together with the required selection rules for the transitions. Therefore,
in a sense, without trying to oversimplify, one would expect 
the matching procedure between the two methods (inner region/basis
representation - outer region/grid representation of the wavefunction)
to hold regardless of the actual system being multi-electron or
single-electron in nature. It is for this reason we believe the formulation in the
present work should be readily extendable to complex multi-electron systems.
The theoretical details and subsequent application will be
more complicated, due to the multiplicity of ionizing channels for the
ejected electron in such cases.
In the following sections we will develop our approach for the one-electron
atom case in detail thereby laying bare the basic concepts of what we believe
a novel combination of basis set and finite-difference methods.
\begin{figure}
\centering \includegraphics[scale=0.3]{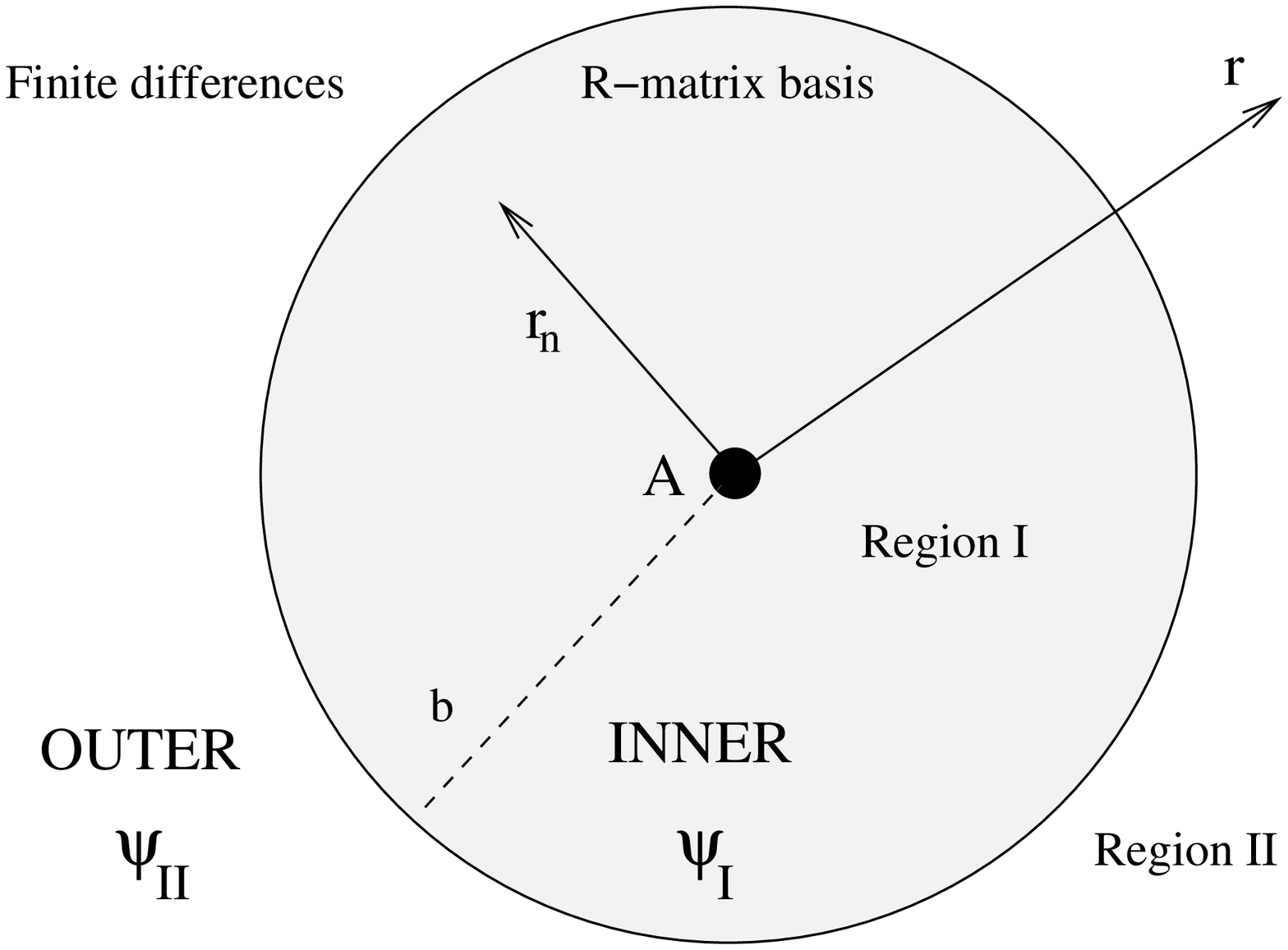} 
\centering \includegraphics[scale=0.3]{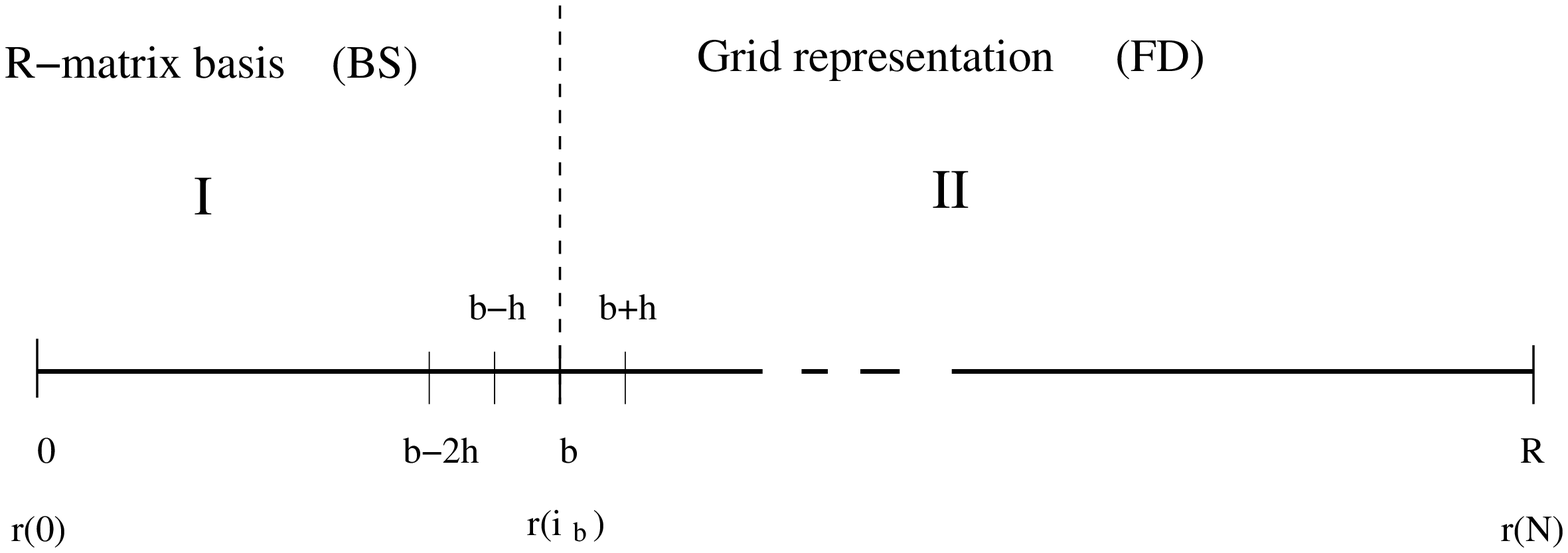}
\caption{Partition of configuration space for the electron coordinate.
In the inner region $I$ an eigenstate expansion representation
of the wavefunction is chosen, while in the outer region $II$
a grid representation is considered.}
\label{fig:fig0} 
\end{figure}

\section{The theoretical framework} 
\label{sec:345}
In this section we develop a theory for solving the TDSE using basis and grid representations in the inner and outer 
regions, respectively. The artificial R-matrix division of configuration space into two regions causes 
time-dependent boundary terms to appear in the corresponding TDSEs (in regions $I$ and $II$) 
which exactly account for the amount of probability current passing through the boundary during the interaction with 
the external field as well as after its turn off. Since the time-dependent wavefunction consists of two parts,  a 
careful analysis is necessary in order to obtain the physical observables of interest such as bound and ionization probabilities 
as well as energy and angular information on the ejected electron. 

In Sub-Sec.~\ref{sec:3} we present the calculation of the R-matrix
eigenstates defined in region $I$ and derive the time-evolution
equations for a wavefunction expanded over the R-matrix eigenstates
of the field-free Hamiltonian. In Sub-Sec. ~\ref{sec:4} we derive the finite-difference TDSE 
governing the radial motion of the ejected electron. In Sub-Sec. ~\ref{sec:4.5} we summarize the calculational 
procedure for the forward in time propagation of the wavefunction. Finally, in Sub-Sec.~\ref{sec:5} we give the 
formal expressions for the calculation of experimental observables adapted to our methodology.

Before proceeding further we first define the inner and outer
region as shown in Fig. \ref{fig:fig0}. In region $I$ (defined as $[0,b]$) the TD wavefunction
$\psi_{I}$ is expanded over the eigenstates of the Hamiltonian matrix representation 
in the interval $[0,b]$. In region $II$ (defined as the interval
$[b,R]$) the TD wavefunction is represented by its values
[$\psi_{II}({\bf r}_{i},t)$] at equidistant grid points $r(i)=ih,i=i_{b},i_{b}+1,...,N$.

\subsection{R-matrix basis-set TDSE in the inner region}
\label{sec:3} 
In the inner region $I$ we define the radial channel functions
$\bar{f}_{l}(r)$ which are expanded over the R-matrix basis set $P_{kl}(r)$ (
defined in appendix \ref{sec:a1}) as:
 \begin{equation}
\bar{f}_{l}(r,t)\equiv\sum_{k=1}^{K}C_{kl}(t)P_{kl}(r),\qquad0\le r\le b,
\label{eq:fl_pl}
\end{equation}
 with the bar at the top indicating that the channel function has
been obtained by summing over the radial Hamiltonian eigenstates of
the \textit{inner region}. Note that we have ignored the dependence on the magnetic and spin quantum 
numbers. From the above definition and the TD wavefunction [Eq. (\ref{eq:wf_lm})] we obtain in the inner region $I$:
 \begin{equation}
\psi_{I}({\bf r},t)=\sum_{k=1}^{K}\sum_{l=0}^{L}C_{kl}(t)\frac{P_{kl}(r)}{r}Y_{l0}(\theta,\phi), \qquad0\le r\le b.
\label{eq:wf_bs}
\end{equation}
The time evolution of the TD wavefunction is now entirely contained
in the coefficients $C_{kl}(t)$. 
The time evolution of the $C_{kl}(t)$ is determined by the TDSE.  However in
writing the TDSE we must take care that the Hamiltonian and dipole operators which act on  $\psi_{I}({\bf r},t)$ are
Hermitian over region $I$ (where $\psi_{I}({\bf r},t)$ is only defined).  
The Hermitian inner-region Hamiltonian is given by $H_I = H_0 + \hat{L}_h$ and  the dipole operator by
$D_I = D + \hat{L}_d$, where the Bloch surface terms $\hat{L}_h$ and $\hat{L}_d$ are set out in Eqs. (\ref{eq:bloch-h}) and (\ref{eq:bloch-d}) respectively.  In these circumstances the TDSE over region $I$ is written:
 \begin{equation}
i\frac{d\psi_{I}}{dt}({\bf r},t)=\left[H_{I}+D_{I}(t)\right]\psi_{I}({\bf r},t)-\left[\hat{L}_{h}+\hat{L}_{d}(t)\right]\psi({\bf r},t),
\label{eq:tdse_i}
\end{equation}
with $0 \le r \le b$.
This equation is a key one to the method.  The second term on the right hand side compensates for the Bloch terms introduced to make $H_{I}$ and $D_{I}$ Hermitian.  
Note that it makes a contribution only at $r=b$ and brings into play there
$\psi({\bf r},t)$ {[}Eq. (\ref{eq:wf_lm})], a wavefunction form which we have defined throughout both regions.  
This term is central to any time propagation scheme in region $I$ because it connects the wavefunction form
$\psi_{I}({\bf r},t)$ specific to that region (which may be multi-electronic in a more general formulation) with a wavefunction form that at $r=b$ represents a single electron and which in calculations is obtained from region $II$. 
We obtain from Eq.(\ref{eq:tdse_i}) the evolution equations for the coefficients $C_{kl}(t)$ by projection over the
states $(P_{kl}(r)/r)Y_{l0}(\theta,\phi)$: 
\begin{eqnarray}
i\frac{d}{dt}C_{kl}(t) & = & \sum_{k^{\prime}l^{\prime}}\left[\epsilon_{k^{\prime}l^{\prime}}\delta_{kk^{\prime}}\delta_{ll^{\prime}}
+D_{kl,k^{\prime}l^{\prime}}(t)\right]C_{k^{\prime}l^{\prime}}(t)
\nonumber \\
&-&\frac{1}{2}P_{kl}(b)F_{l}^{\prime}(b,t). \nonumber
\end{eqnarray}
The quantity $F_{l}^{\prime}(b,t)$ is defined as: 
\begin{equation}
F_{l}^{\prime}(b,t) = \frac{df_l(b,t)}{dr} - i\frac{A(t)}{c}  \sum_{l^\prime = l \pm 1}  K_{ll^\prime} f_{l^\prime}(b,t),
\label{eq:df_b}
\end{equation}
where $A(t)$ is the time-dependent field potential in the Coulomb gauge (see appendix \ref{sec:a0}) and $K_{ll^\prime}$ is an angular factor given in Eq. (\ref{eq:tdse-potentials_3}). If the coefficient vector ${\bf C}(t)$ is structured as $C^{T}(t)=[C_{10}(t),...,C_{K0},C_{11}(t),...C_{K1},....,C_{1L}(t),...,C_{KL}(t)]$ the 
inner-region TDSE in matrix notation is as follows: 
\begin{equation}
\dot{C}_{kl}(t)=-i[{\bf H}\cdot{\bf C}]_{kl}(t)+\frac{i}{2}w_{kl}F_{l}^{\prime}(b,t).
\label{eq:tdse_inner}
\end{equation}
The amplitudes $w_{kl}$
have been defined as $w_{kl}=P_{kl}(b)$ in appendix \ref{sec:a1}.
The matrix ${\bf H}$ has the block-triangular form of Eq. (\ref{eq:h_matrix})
with the block-diagonal matrices $\hat{h}_{l}$ and the lower- and-upper
block matrices $\hat{D}_{ll^\prime}(t)$ having matrix elements as:
\begin{align}
\langle kl|\tilde{h}_{l}|k^{\prime}l\rangle&=\epsilon_{kl}\delta_{kk^{\prime}},\\
\langle kl|\hat{D}_{ll\pm1}|k^{\prime}l\pm1\rangle &=-i\frac{A(t)}{c}K_{ll\pm1}\tilde{t}_{kl;k^{\prime}l\pm1}(r),
\label{eq:inner_evolution}
\end{align}
where $\tilde{t}_{kl;k^{\prime}l\pm1}(r)$ are matrix elements defined in Eq. (\ref{eq:dv_hermitian}). 

\subsection{Finite-difference TDSE in the outer region}
\label{sec:4} 
In the external region $II$ a grid representation of the TD wavefuction is adopted: 
\begin{equation}
\psi_{II}({\bf r}(i),t)=\sum_{l=0}^{L}\frac{f_{l}(i,t)}{r(i)}Y_{l0}(\hat{r}),\qquad b\le r(i),
\label{eq:wf_fd}
\end{equation}
with $i=i_{b},..,I$. The time-dependence of the wavefunction is represented by the
values of the radial channel functions on a equidistant discretized
grid, $f_{l}(i,t)=f_{l}(r(i),t)$ with $h=r(i+1)-r(i),i=i_{b},..,I$.
The grid is defined such that $r(i_{b})=b$ and $r(I)=R$. Furthermore
by constructing the vector ${\bf F}(t)$ from the values of the radial
channels $f_{l}(i,t)$ at the grid points, we obtain a vector of length
$L\times I$ structured as ${\bf  F}^{T}(t)=\left[f_{0}(i_{b},t),...,f_{0}(I,t),f_{1}(i_{b},t),...,f_{1}(I,t),...,f_{L}(i_{b},t),...f_{L}(I,t)\right]$.
The FD representation of the TDSE takes the form:
\begin{equation}
\dot{f}_{l}(i,t)=-i [{\bf H}\cdot{\bf F}]_{l}(i,t).
\label{eq:tdse_matrix_fd_0}
\end{equation}
In the FD representation of the time-dependent Hamiltonian ${\bf H}(i,t)$ the entries
$\hat{h}_{l}$ and $\hat{D}_{ll\pm1}(t)$ are square matrices of order
$I-i_{b}+1$. The explicit form of these operators depends on the
approximation chosen for the derivatives. In the present case, the
first and the second derivative of a function $\phi(r)$ are approximated
with a 5-point central difference scheme as follows: 
\begin{equation}
\frac{d^{q}}{dr^{q}}\phi(i,t)=\sum_{j=-2}^{2}\frac{d_{j}^{(q)}}{h^{q}}\phi(i+j,t),\quad q=1,2,
\label{eq:deriv}
\end{equation}
 with $d_{j}^{(q)}$ chosen so that polynomials of order 4 are differentiated 
exactly. Given the above, the finite-difference approximation of the
diagonal operator in the FD Hamiltonian is, 
\begin{equation}
\hat{h}_{l}f_{l}(i,t)=-\frac{1}{2}\sum_{j=-2}^{2}\frac{d_{j}^{(2)}}{h^{2}}f_{l}(i+j,t)
+\left[ \frac{l(l+1)}{2[r(i)]^{2}} +V(i) \right] f_{l}(i,t).
\label{eq:fd_diagonal_operator}
\end{equation}
The velocity form of the non-diagonal operator is given by: 

\begin{eqnarray*}
\label{eq:fd_non_diagonal_operator_v}
\hat{D}_{ll\pm1}(t)f_{l\pm1}(i,t) 
=&
\frac{-iA(t)}{c}\!
\left[ 
\sum_{j=-2}^{2}\frac{d_{j}^{(1)}}{h}f_{l\pm1}(i+j,t) 
\right.  \\ 
 & \left. 
-(l-l^{\prime})\frac{l_{>}}{r(i)}f_{l\pm1}(i,t)
\right] K_{ll\pm1}. \nonumber
\end{eqnarray*}

The FD  form of the TDSE  in Eq. (\ref{eq:tdse_matrix_fd_0}) is sufficient to propagate
the wavefunction in time provided it vanishes at both ends
of the spatial grid at all times.  This is certainly the case when the FD grid has its
innermost point at the origin.  
In contrast, in the present case, vanishing boundary conditions occur only at the
far end of the grid ($r=R$). More specifically, we assume
that $f_{l}(I-1,t)=f_{l}(I,t)=0$ for all $l$ and this forms the
set of boundary conditions imposed on the wavefunction at the far
boundary.

Thus some further consideration of the differential operators involved
in the FD representation of the TDSE is necessary and we shall shortly see that non-zero
function values at an inner boundary $r=r(i_b)=b$ bring about contributions from
functions values at points below the inner boundary point to the propagation. 

We begin by appreciating that since the FD method
is a local method, the evaluation of function derivatives at any point
relies on function values at neighbouring points, and which of these
come into play depends on the approximation chosen for the derivatives,
as mentioned earlier.  In the FD method the operators are also discretized in a
similar way to the functions, i.e. as $\hat{O}(r,t)=\hat{O}(i,t)$.
The action of a non-derivative operator on a function is trivial,
since $\hat{O}(r)\phi(r)=\hat{O}(i)f(i)$ at the $i-$th grid point,
but the same is no longer true when operators contain derivatives.
Then the rule of differentiation should be given. The central characteristic
of the differential operators in the FD method is that values of the
wavefunctions at neighbouring points are involved in the calculation
of the derivative function. It is then obvious that since the diagonal
operators in the finite-difference TDSE {[}Eq. (\ref{eq:fd_diagonal_operator})]
involve the second-order differential operator (due to the kinetic
term) the complete determination of the ${\bf H}\cdot{\bf F}$
requires knowledge of the $f_{l}(i,t)$ at points $i=i_b-1,i_b-2$
since these enter the determination of second-order derivatives at
points $i_b$ and $i_{b}+1$ according to Eq. (\ref{eq:deriv}). If the propagation is
done in the velocity gauge a similar conclusion
is reached by considering the non-diagonal operators $\hat{D}_{ll^{\prime}}(t)$. 
The modified form of the TDSE corresponding to a non-vanishing solution on the inner boundary is then
\begin{eqnarray}
\label{eq:tdse-outer}
\dot{f}_{l}(i,t)&=&-i[{\bf H}\cdot{\bf F}]_{l}(i,t) \\
&+&\delta_{ii_{b}}\left[B_{0l}(i_{b}-1,t)+B_{0l}(i_{b}-2,t)\right]  \nonumber\\
 &+&\delta_{ii_{b}+1}B_{1l}(i_{b}-1,t),                                                             \nonumber
\end{eqnarray}
where
\begin{subequations} 
\label{eq:bc_ii}
\begin{align}
B_{1l}(i_{b}-1,t) & =-\frac{d_{-2}^{(2)}}{2h^{2}}\bar{f}_{l}(i_{b}-1,t)+\frac{d_{-2}^{(1)}}{h}\bar{g}_{l}(i_{b}-1,t)
\label{eq:bc_ii_1}\\
B_{0l}(i_{b}-1,t), & =-\frac{d_{-1}^{(2)}}{2h^{2}}\bar{f}_{l}(i_{b}-1,t)+\frac{d_{-1}^{(1)}}{h}\bar{g}_{l}(i_{b}-1,t)
\label{eq:bc_ii_2}\\
B_{0l}(i_{b}-2,t), & =-\frac{d_{-2}^{(2)}}{2h^{2}}\bar{f}_{l}(i_{b}-2,t)+\frac{d_{-2}^{(1)}}{h}\bar{g}_{l}(i_{b}-2,t)
\label{eq:bc_ii_3}
\end{align}
 \end{subequations} 
and $\bar{g}_{l}(r,t)$ are given by,
\begin{equation}
\bar{g}_{l}(i,t)=-i\frac{A(t)}{c}
\left[
K_{(l-1)l}\bar{f}_{l-1}(i,t)
+
K_{l(l+1)}\bar{f}_{l+1}(i,t)
\right].\nonumber
\end{equation}
The elements $K_{ll^\prime}$ are given by Eq.(\ref{eq:tdse-potentials_3}) but when $l=L$ the term 
with $K_{l(l+1)}$ is missing and when $l=0$ the term with $K_{(l-1)l}$ is also missing from the corresponding
equations. The bar on the $\bar{f}_{l},\bar{g}_{l}$ emphasizes that these radial function values have been
evaluated by use of the R-matrix basis set expansion form of the
wavefunction in region $I$.

Eq. (\ref{eq:tdse-outer}) is the second (and last!)  key equation of the method. 
It does for region $II$ what Eq. (\ref{eq:tdse_i}) above did for region $II$.  The communication with the solution in region $I$  is provided through the terms involving radial function evaluations at two FD points in region $I$ immediately
inside the boundary with region $II$.  Although our detailed exposition above has centred around one-electron wavefunctions throughout both regions, it is clear how the concept embodied in Eq. (\ref{eq:tdse-outer})  can be extended to handle a region $I$ that is multi-electron in character.  The crucial requirement of such a multi-electron inner region is that it must collapse to one-electron character within a few FD points of its outer boundary at $r=b$.  Since in multi-electron R-matrix calculations anyway the inner region must be one-electron in nature by $r=b$,  our additional requirement provides no great extra overhead.

\subsection{Calculational procedure} 
\label{sec:4.5}

Having set out the form of the TDSE in the two regions $I$ 
 [Eq. (\ref{eq:tdse_inner})] and $II$ [Eq. (\ref{eq:tdse-outer})]  we now
present briefly the computational procedure involved in the propagation of the 
wavefunction $\psi({\bf r}, t)$ through one time-step from time $t$ to time $t+ \tau)$. 
\paragraph{Outer region: calculation of $\psi_{II}({\bf r}, t + \tau)$:}
Assuming at time $t$ the wavefunction is known throughout the inner and outer regions $I,II$ we first consider the 
outer region $II$ TDSE [Eq. (\ref{eq:tdse-outer})].  Although there is a wide variety of methods in the literature we have chosen to employ the standard Taylor propagator as prescribed in  Eq. (\ref{eq:taylor_sum}). The evaluation of the Taylor series terms requires the quantities $B^{(0)}_{1l}(i_{b}-1,t), 
 B^{(0)}_{0l}(i_{b}-1,t),B^{(0)}_{0l}(i_{b}-2,t)$ which bring into play values of the partial waves
 $\bar{f}_l(i-2,t),\bar{f}(i-1,t)$ evaluated \emph{in the internal region} 
at time $t $ [Eq. (\ref{eq:bc_ii})]. These inner-region partial wave values are formed using Eq. (\ref{eq:fl_pl}). 

\paragraph{Inner region: calculation of $\psi_I({\bf r}, t + \tau)$:}
In a similar way as done for the outer region, the propagation of the coefficients $C_{kl}(t)$ from time $t$ through one time-step to gain their values $C_{kl}(t+\tau)$
at time $t+\tau$ is now based on the inner-region TDSE in the form of Eq. (\ref{eq:tdse_inner}) and the Taylor expansion Eq. (\ref{eq:taylor_sum}). For this evaluation 
knowledge of the quantity $F_l^\prime(b,t), l=0,1,..,L$ at time $t$ is required. The latter quantity includes the 
outer-region partial wave $f_l(b,t)$ and its derivative $f_l^\prime(b,t)$ evaluated \emph{on the boundary $r=b$}. 
Having calculated the coefficients $C_{kl}(t+\tau)$  we can immediately form the wave function 
$\psi_I({\bf r},t+\tau)$ according to Eq. (\ref{eq:wf_bs}). 

By this stage the wavefunction is known at time $t+\tau$ throughout regions $I$ and $II$ and we can
proceed further in time by repeating the above procedure for successive time-steps $\tau$.

\subsection{Observables within the dual representation}
\label{sec:5}
In this section we develop the necessary formulation for the calculation of observables given the different representation 
used of the time-dependent wavefunction in the inner and outer region (regions $I$ and $II$ respectively). 
These representations are given by Eq. (\ref{eq:wf_bs}) and Eq. (\ref{eq:wf_fd}), respectively. Any spatially dependent observable represented by the operator 
$\hat{O}({\bf r},t)$ is calculated through the standard formula,
$O(t) = \langle \psi({\bf r},t )|\hat{O}({\bf r}, t) | \psi({\bf r},t)\rangle $ 
which in our case separates into two pieces. To link with the standard experimental setups we assume that any calculation of the observables is performed 
for times where the external field has vanished. In the following formulas, taking the pulse duration as $T$, we
assume the projection time $t_p$ such that $t_p \ge T$.
To obtain the population $W_{nl}(t_p)$ in an eigenstate of the physical system
 $\phi_{nl}({\bf r})= (F_{nl}(r)/r)Y_{l0}(\hat{r})$ at time $t_p$,  we use the projection operator
$\hat{P}_{nl}= |\phi_{nl} \rangle  \langle \phi_{nl}|$ with the result:
\begin{equation}
 W_{nl}(t_p) = \left| ( F_{nl}|\bar{f}_l )_{I} +  ( F_{nl}| f_l )_{II} \right|^2,
\label{eq:population}
\end{equation}
with $\bar{f}_l(r,t)$ given by Eq.~(\ref{eq:fl_pl}) and $(a|b)_I, (a|b)_{II}$ denoting radial integrations over the
inner and outer regions, respectively.
Complete information about the final state (ignoring spin variables) is possible by recalling the partial wave expansion of a
continuum electron with asymptotic momentum ${\bf k}=(k,\theta_{k}, \phi_{k})$, namely:
\begin{equation}
\psi^{(-)}_{{\bf k}}({\bf r}) = \sum_{lm_l} a_{lm_l}(k) \frac{1}{r}F_{kl}(r)Y^{\star}_{lm_l}(\hat{k})Y_{lm_l}(\hat{r}), 
\end{equation}
where $\hat{k}=(\theta_{k},\phi_{k})$ defines the direction of the photoelectron with respect to the polarization axis 
(quantization axis), $F_{kl}(r)$ is normalized on the energy scale and the amplitudes 
$a_{lm_l}(k)$ are chosen so that the wavefunction $\psi^{(-)}_{{\bf k}}({\bf r})$ fulfils incoming spherical wave
boundary conditions. In the present case, where the ionizing target is hydrogen and $m_l = 0$ 
(in the following again we drop the $m_l$ dependence) we have 
$a_l(k) = i^{l}e^{-i\sigma_l(k)}$ with $\sigma_l(k)$ the long-range Coulomb phase shift analytically known 
\cite{messiah:1999}. 
Therefore the desired angular distribution is obtained through the projection operator 
$\hat{P}_{k} =  |\phi^{(-)}_{\bf{k}} \rangle  \langle \phi^{(-)}_{{\bf k}}|$ which gives:
\begin{displaymath}
\frac{dW(\epsilon_k, \hat{k}, t_p)}{d{\bf k}} =\left| \sum_{l}
\left[ (F_{kl}|\bar{f}_l)_{I} + (F_{kl}|f_l)_{II}  \right] a_l(k) Y_{l0}(\hat{k})
\right|^2,
\label{eq:pad_pes}
\end{displaymath}
with $d{\bf k} = k^2 dk d\Omega_k$ the volume element in momentum space.
Integration of the above formula over the kinetic energies  $\epsilon_k$ ($\epsilon_k = k^2/2$) 
results in the photoelectron angular distribution (PAD),
\begin{equation}
 \frac{dW(\epsilon_k, t_p)}{d\Omega_k} = \int dk k^2 \frac{dW(\epsilon_k, \hat{k}, t_p) } {d{\bf k}},
\label{eq:pad}
\end{equation}
while integration over the photoelectron ejection angles $(\theta_k, \phi_k)$  provides the angle-integrated 
photoelectron energy distribution (PES), 
\begin{equation}
\frac{dW(\epsilon_k, t_p)}{d\epsilon_k} = \sum_{l}
\left| (F_{kl}|\bar{f}_l)_{I} + (F_{kl}|f_l)_{II}  \right|^2_{k = \sqrt{2\epsilon_k}}.
\label{eq:pes}
\end{equation}
Finally, further integration over the photoelectron kinetic energies of the last equation results in the total ionization 
probability (yield) at time $t_p$ as:
\begin{equation}
W(t_p) = \int d \epsilon_k \frac{dW(\epsilon_k, t)}{d\epsilon_k}.
\label{eq:yield}
\end{equation}
 At this point we have completed the present theoretical formulation leading to the calculation of
the most important experimental observables following the interaction of an electromagnetic field with a one-electron
atomic target in the dipole approximation.


\section{Illustrative application to hydrogen}
\label{sec:6}
In the present section we apply our approach to the case of ionization
of the hydrogen atom by a strong EM field. The reasons we have chosen hydrogen are as follows:
 (a) it represents the simplest among the atomic systems
having just one electron participating in the ionization process, thus being
free from complications that may arise from inter-electronic effects
in the case of multi-electron systems, (b)  angular momentum
considerations are reduced to the minimum level where a simple partial
wave expansion is adequate to represent the TD wavefunction throughout
the electron's configuration space  and (c) last but not least very reliable methods 
treating one-electron systems \cite{2001:8,2007:37,taylor:2005}
are at our disposal for a systematic study of the reliability and accuracy issues surrounding the present method. 
In the present application we have chosen an explicit type time-propagator based on a Taylor 
expansion [Eq. (\ref{eq:explicit})]. In all the calculations the order of the 
propagator was $P=12$ and the time step  $\tau=1.5625\times10^{-4}$ a.u. .

\subsection{Initial state calculation}
\begin{figure}
\centering\includegraphics[scale=0.3,angle=270]{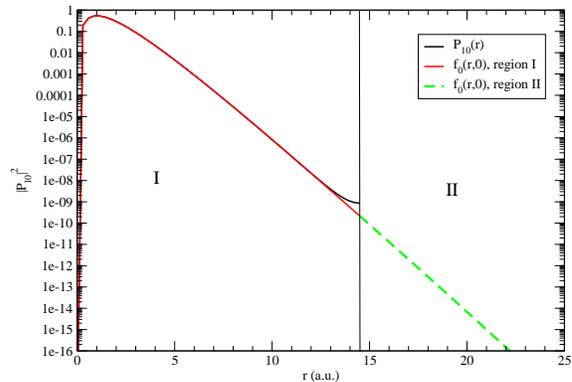} 
\caption{(Color online) Initial state wavefunction calculated by an imaginary field-free time propagation.
The spatial step was $h=0.29$ a.u. and the time step was $\tau =1.5625\times10^{-4}$
a.u. The order of the Taylor propagator was 12. Black curve represents the trial wavefunction while the red (solid) and the 
green (dashed) curves represent the converged wavefunction for the inner and outer regions, respectively.}
\label{fig:fig1} 
\end{figure}

We start by calculating the $P_{10}(r),\,\,0\le r\,\le b$
radial function by numerically solving the radial SE for $l=0$ [Eq. (\ref{eq:h0})] within the
inner region. The initial state, made up of $\psi_{I}(\mathbf{r},t=0)$
and $\psi_{II}(\mathbf{r},t=0)$ in the inner and outer region, respectively,
is then calculated by means of an imaginary time propagation of the field-free versions
of  Eqs.  (\ref{eq:tdse_inner}) and (\ref{eq:tdse-outer}) with initial conditions: 
\begin{eqnarray*}
C_{kl}(t=0) & = & \delta_{kl;10},\\
f_{l}(i,t=0) & = & \delta_{l0}\delta_{ii_b}P_{10}(b).
\end{eqnarray*}
It is important to emphasize here, that the R-matrix eigenstates do not actually represent the eigenstates 
of the system, instead they only serve as a complete basis for the representation of the physical state exclusively
in the interval $[0,R]$.  The B-splines basis used consisted of $n_{b}=57$ basis functions of order
$k_{b}=9$.  The knot-sequence is chosen to be linear with discretization step equal to that of the outer
region spatial step $h=0.29$ a.u.. 
In Fig. \ref{fig:fig1} we plot the squared amplitude of the radial part of the R-matrix basis 
for $n=1,l=0$, $P_{10}(r)$ (black curve) and the state 
$P_{1s}(r,t \rightarrow \infty )=r\langle Y_{10}|\psi(\mathbf{r},t \rightarrow i\infty) \rangle,\: 0\le r\le R$ 
as converged after the imaginary-time field-free propagation. Black-solid and red-dashed curves represent
the inner ($I$) and outer ($II$) region values.  The inner boundary has been set at $b=14.5$ a.u. while the outer 
boundary at $R=174$ a.u. The R-matrix eigenstate $P_{10}(r)$ has zero-derivative on the boundary 
(due to the chosen boundary condition $P_{kl}^{\prime}(b)=0$) while the imaginary time propagation has converged
to the state $P_{1s}(r)$ with non-vanishing derivative on the boundary as actually is the case for the ground state 
of hydrogen. Our initial state has the following form:
\begin{equation}
 \psi({\bf r},0) = \frac{1}{r}Y_{10}(\hat{r})\left\{ 
\begin{array}{ll}
\sum_{k} C_{k0}( 0 ) P_{k0}(r),  &  \mbox{region } I \\
f_0(i,0),               &  \mbox{region } II
\end{array}
\right. .
\label{eq:initial_state}
\end{equation}

\subsection{Real time propagation}
Having obtained an accurate initial state [Eq. (\ref{eq:initial_state})], through imaginary time
propagation, we proceed to the propagation of the TDSE in the outer/inner region in the presence of an external EM field.
The EM field chosen was linearly polarized along the $z$-axis with vector potential: 
\begin{equation}
A(t)=A_{0}\sin^{2}(\frac{\pi}{T}t)\sin\omega t,
\label{eq:a_field}
\end{equation}
where $\omega=2\pi/T_{0}$ is the field frequency and $T=nT_{0}$
the pulse duration ($n$ being the number of cycles contained in the
pulse and $T_0$ the field period). 
\begin{figure}
\centering \includegraphics[scale=0.3,angle=270]{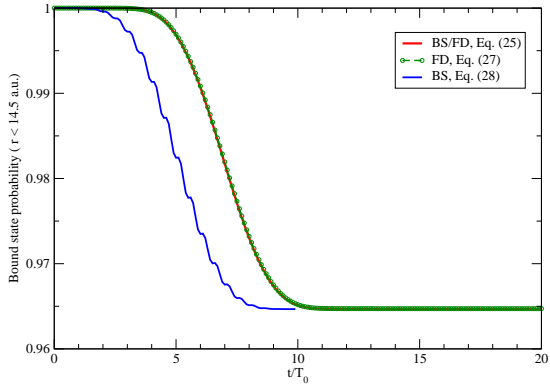}
\caption{(Color online) Hydrogen bound state population within the $(0,14.5)$ a.u. region after irradiation by an external EM field (see text for the field details). Curves represent the present mixed method (BS/FD) as well as standard 
 finite-difference (FD) and eigenstate expansion (BS) methods.}
\label{fig:fig2} 
\end{figure}
To test our approach we calculated the bound state population using three different methods. First, within the present 
method (BS/FD), we obtained the population in the $[0,b]$ region by simply summing 
(at a sufficiently long time $t_p$) over all R-matrix eigenstates as:
 \begin{equation}
P^I_b(t_p)=\sum_{kl}|C_{kl}(t_p)|^{2}.
\label{eq:pop_b_bf}
\end{equation}
The second method consisted of the standard finite-difference (FD) approach over both regions $I$ and $II$ i.e. over 
the whole range $[0,R]$ (with the same spatial and time step) thereby invoking no division of the electron's configuration space.  Formally, within the present method this is equivalent to setting $b=0$. We calculated the ionization probability $P_c$ as:
\begin{equation}
P_{c}(t_p)= \sum^{r(i) \ge b}_{il}|f_{l}(i,t_p)|^{2},
\label{eq:pop_c_fd}
\end{equation} 
where $t_p$ was chosen sufficiently large so that all the outgoing components of the ionized wavepacket were 
able to travel beyond the chosen distance $b$. With no absorbing potential present we always have for the bound state 
probability $P_b(t) =1-P_c(t)$.  When an absorbing potential is present then the bound state probability is obtained as:
\begin{equation}
P_{b}(t_p)= \sum^{r(i) \le b}_{il}|f_{l}(i,t_p)|^{2}.
\label{eq:pop_b_fd}
\end{equation} 
Finally, we performed calculations using a standard basis set (BS) to span the whole range $[0,R]$ with again no
division of configuration space. This is formally equivalent to setting $b=R$. We obtained the bound state population by summing only over the bound part of the spectrum:
\begin{equation}
P^{(BS)}_b(t_p)=\sum_{kl}|C_{kl}(t_p)|^{2}, \qquad \epsilon_{kl} \le 0.
\label{eq:pop_b_bs}
\end{equation}
\begin{figure}
\centering \includegraphics[scale=0.3,angle=270]{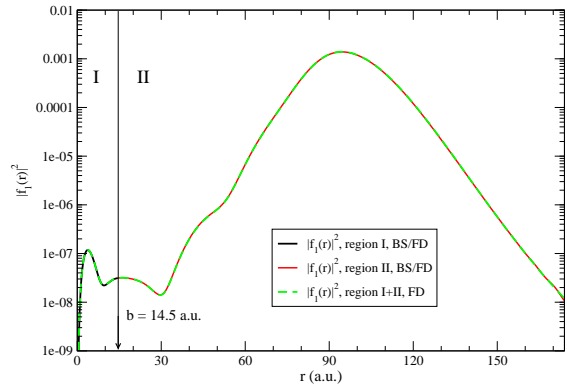}
\caption{(Color online) Absolute square of the hydrogen partial wave $l=1$ ($|f_l(r,t_p)|^2$) at $t_p = 157.1 $ a.u. 
after irradiation by an external EM field (see text for the field details). Curves represent results with 
the present mixed method (BS/FD) as well as with the standard finite-difference (FD) method.}
\label{fig:fig2-1} 
\end{figure}
In Fig. \ref{fig:fig2} we show the bound state population of hydrogen as a function of time when the atom is irradiated 
by a pulse of central frequency $\omega = 0.8$~a.u. ($21.769$ eV), total duration of 10 cycles ($T=10T_0 = 10\times 7.854 = 78.54 $ a.u.) and peak intensity $I_0 = 10^{14}$ W/cm$^2$. 
For the BS calculation the bound state population is calculated at the end of the pulse (10 cycles) and no further
field-free propagation of the wavefunction is required since the population distribution remains unchanged.
In the case of the FD and BS/FD calculations the propagation is extended for a further
10 cycles  (field-free propagation) after the end of the pulse until a sufficiently large part of the wavepacket has
passed the artificial boundary at $r= b=14.5$ a.u.. Given the photon frequency, the hydrogen ionization potential and the 
rather modest field intensity, we expect the dominant partial wave in the outer region to be the $l=1$ partial wave with the 
 electron's kinetic energy peaked around $\epsilon_k \sim 0.8-0.5=0.3$ a.u. (8.16 eV). By assuming an outgoing wavepacket
with central energy of  $0.3$ a.u. (thus of velocity $k = 0.7746$ a.u.) we can estimate that 10 cycles of field-free 
propagation is sufficient for our purposes. In connection with this latter point note that this wavepacket travels 
a distance of 14.5 a.u. in approximately 2.5 field-cycles. This is why the FD and BS/FD bound state populations exhibit a time-delay 
compared to the BS bound state population. The maximum angular momenta allowed was $L = 3$. Results remained practically unchanged 
against further increase in angular momentum. We have performed similar calculations with peak intensities $I_0=10^{15}$ 
W/cm$^{2}$ and found similar results with analogous agreement between the BS/FD, FD and BS bound state populations.

In Fig. \ref{fig:fig2-1} the values of $|f_l(r,t_p)|^2, l=1$ are plotted as calculated with the present BS/FD and the 
standard FD method at $t_p = 20 T_0 = 157.1 $ a.u.. In region $I$ (within 14.5 a.u.  of the nucleus) the partial wave 
function ($|\bar{f}_1(r,t_p)|^2$) was obtained from Eq. (\ref{eq:fl_pl}). In region $II$ (from 14.5 a.u out to 174 a.u.) 
the values of $|f_1(r,t_p)|^2$ come directly from the propagation of  the outer-region TDSE [Eq. (\ref{eq:tdse-outer})]. Similarly for the FD 
calculation we obtained $|f_1(r,t_p)|^2$ by solving Eq. (\ref{eq:tdse_matrix_fd_0}) over the whole range $[0,R]$. The figure 
displays excellent agreement  between such results from the present (BS/FD) method and the 
standard FD method. We have chosen to plot only the $l=1$ partial wave since this is the dominant outgoing 
channel with all other partial wave channels being an order of magnitude lower. This observation simplifies the analysis
of the physics involved in the process. We briefly elaborate on this plot.  The peak probability for 
the travelling wavepacket appears around $\sim 92 $ a.u. with a much smaller secondary peak inside
region $I$. In an energy representation of the wavepacket, the large peak is associated with the continuum states contribution while 
the second peak is related to the bound states contribution. Whereas the bound contribution is trapped in the inner region  the outgoing component (corresponding to the continuum spectrum) travels a distance of about $r \sim v \times 15T_0 =  0.7746 \times 15 T_0 \sim 91 $ a.u. 
which is rather close to the maximum of the wavepacket probability in the plot. We have allowed 15 cycles of travelling time 
for the wavepacket since significant ionization only takes place around the maximum of the 
applied pulse which occurs at approximately 5 cycles after the turn-on. 
\begin{figure}
\centering \includegraphics[scale=0.3,angle=270]{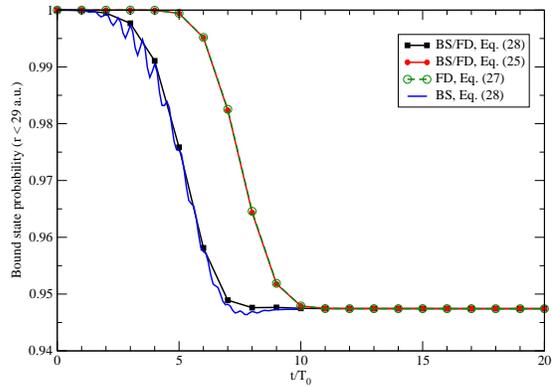}
\caption{Hydrogen bound state population within the $(0,29)$ a.u. region after irradiation by an external EM field 
 of photon frequency of $\omega=0.35$ a.u. and peak intensity $I=10^{14}$ W/cm$^{2}$. Curves represent the 
present mixed method (BS/FD) as well as standard finite-difference (FD) and eigenstate expansion (BS) methods.}
\label{fig:fig3} 
\end{figure}

In Fig. \ref{fig:fig3} the bound state population of hydrogen is shown after exposure to an EM field of central
frequency $\omega = 0.35$~a.u. ($9.524$ eV), total duration 10 cycles and peak intensity
 $I_0 = 10^{14}$ W/cm$^2$. Since the photon energy is comparable to the energy gap ($\sim 10.277$ eV) between the ground and first excited states
($2{\rm s},2{\rm p}$) an appreciable population in these excited states appears at the end of
the pulse.   At the end of the pulse we obtained from the BS calculation a value $P_{g}=0.7368$ for the ground state probability ;
a value $P_{e}=0.2104$ for the total population in all the excited states  ($\epsilon_{kl} < 0 $ ) and  thus
a total bound state probability of $P^{(BS)}_{b}=0.947318$  [Eq. (\ref{eq:pop_b_bs})]. 
The bound state probability as a function of time is shown in the figure (blue line).
We  have also performed a FD calculation (with no absorbing boundary present) and calculated the probability within the 
region $[29, R]$ a.u. using Eq. (\ref{eq:pop_c_fd}) and $P_b = 1 - P_c$.  We chose a box with  $R= 522 $ a.u. to prevent 
reflection of the wavepacket at the outer boundary 
over the time interval of interest. In this case the calculated bound state probability for the FD method is 
given by the green curve (empty cycles).  Next, we applied the present method (BS/FD) for $b=29$ a.u. and $R=552$ a.u. 
To maintain the same accuracy in the calculations in the inner region we increased the number of  B-splines basis members to $n_b=108$. 
To compare with the BS and FD calculations we obtained the various probabilities as follows: Black curve (filled squares) 
in the figure was calculated using Eq.  (\ref{eq:pop_b_bf}) which includes a summation only over those R-matrix eigenstates that 
have negative energies such that $\epsilon_k \le 0$ [equivalent to Eq. \ref{eq:pop_b_bs})]. This curve follows closely the 
bound state probability calculated using the BS method. If in Eq. (\ref{eq:pop_b_bf}) we include all 
R-matrix states  then the probability enclosed in region $I$ is given by the red curve (filled circles) and matches perfectly 
with the bound state probability from the FD calculation. Similar evaluation through Eq. (\ref{eq:pop_c_fd}) and 
$P_b = 1 - P_c$ with the BS/FD method results in practically the same curve and verifies the equality of results obtained 
using Eq. (\ref{eq:pop_b_bf}) and Eq. (\ref{eq:pop_c_fd}). In other words any increase/decrease of probability 
within region $I$ is matched by  an equal decrease/increase of probability within region $II$.

\begin{figure}
\centering \includegraphics[scale=0.3,angle=270]{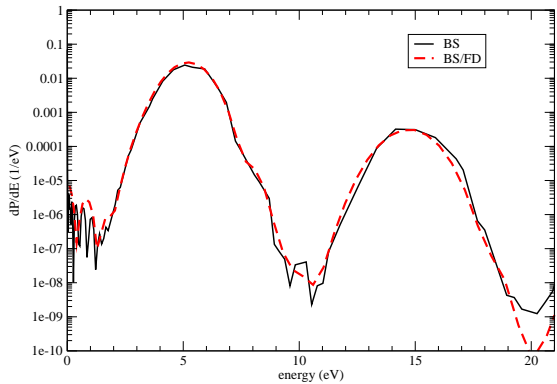} 
\caption{Hydrogen photoelectron energy spectrum after irradiation by an external EM field 
 of photon frequency of $\omega=0.35$ a.u. (9.52 eV) and peak intensity $I=10^{14}$ W/cm$^{2}$. 
Curves represent the present mixed method (BS/FD) and the eigenstate expansion (BS) methods.}
\label{fig:fig3-1} 
\end{figure}

Finally in Fig. \ref{fig:fig3-1} we have calculated the photoelectron energy spectrum up to about 20 eV kinetic energy of the 
ejected electron. In the hydrogen case, although the analytical solutions for the bound and the continuous spectrum  
are available, the numerical calculation of the eigenstates proves more advantageous for the evaluation of the 
necessary integrals. In expression (\ref{eq:pes}) we may use an asymptotic expansion for the Coulomb functions 
$F_{kl}(r)$ \cite{burgess:1963,2007:39} provided that (a) the evaluation is performed at times where the outgoing part 
of the electron wave packet has travelled sufficiently far away from the residual  system (b) the projection operator is constructed  
either from Coulomb wavefunctions or plane waves depending on the chosen projection time ($t_p$) and (c) the inner-region contribution 
is ignored since it is only the bound part of the wavepacket that still remains there as time grows. The results can easily be checked 
by tracing their convergence in time. A detailed discussion of this approach, very well suited to our approach, can be found in ref. \cite{2007:39}. 
The solid black curve represents the result of a BS calculation while the dashed red curve the result of the present calculation.  Had we used a larger box and a finer mesh for the outer region we would be able to calculate even higher (in energy) the corresponding PES for a full comparison with the BS calculation, but this is not the purpose of the present work. 
\section{Conclusion and perspectives}
\label{sec:7}
In conclusion a new \emph{ab initio} time-dependent method for the treatment of the single-electron ionization of  atomic 
and molecular systems under an external electromagnetic field has been set out.  It has been developed in detail for systems that are single-electron throughout
and applied to the simplest case, namely the hydrogen 
atom. The method is based on the division of the configuration space of the ejected electron into two regions $I$ and $II$. 
In region $I$ (which may be multi-electronic) the time-dependent wavefunction is expanded on the basis of R-matrix eigenstates and propagated 
through the time evolution of the expansion coefficients. 
In region $II$ a grid-representation of the time-dependent 
wavefunction is adopted and a finite-difference technique is employed for the representation of the operators. In both 
regions the chosen time-propagator in illustrative calculations is a high-order explicit Taylor propagator. The key point in the present method is 
the time-dependent matching conditions that the inner (region $I$) and outer (region $II$) wavefunctions 
should simultaneously satisfy at each time step. Although these matching conditions have been developed here 
for an explicit time-propagator, the methodology can also be applied for implicit time propagators. The present work 
represents an important step towards the implementation of such a methodology in multi-electron systems 
(atomic and molecular) where the full advantage of the R-matrix technique can be taken into account. 
The straightforward extension of the present approach to the case of a truly multi-electron system is discussed in 
Sec. \ref{sec:2} and is currently the subject of our work. In addition to our fundamental interest in gaining 
an ab-initio description of multi-electron systems under strong laser fields the present work is mainly motivated 
by the development of sources of short wavelength laser light residing well into the VUV or soft XUV regime (HOHG/FEL sources). 
In contrast to long wavelength laser light, the light from such sources tends to interact directly with more than just a single electron
and is able to probe directly the innermost electrons of multi-electrons systems 
thus making the development of new suitable theoretical methods a necessary and formidable task. 

The authors gratefully acknowledge discussions with Dr. Michael Lysaght, Dr. Hugo Van der Hart and Prof. P. G. Burke. The present work has been supported through a  Marie Curie Intra-European Fellowship under contract TDRMX-040766 awarded to LAAN and by the UK Engineering and Physical Sciences Research Council.

\appendix

\section{TDSE of single-active-electron atomic systems over a spherical harmonic basis}
\label{sec:a0} 
The field-free SAE Hamiltonian $H_{0}$ reads, 
\begin{equation}
H_{0}=-\frac{1}{2}\nabla^{2}+V(r),
\label{eq:hydrogen_hamiltonian}
\end{equation}
with the potential $V(r)$ equal to $-Z/r$ for a purely hydrogenic
system (of $Z$ atomic number). Alternatively $V(r)$ could be constructed as a model or Hartree-Fock
potential. The TDSE of the system in an external time-dependent radiation
field, ${\bf E}(t)$ is written as:
 \begin{equation}
i\frac{\partial}{\partial t}\psi({\bf r},t)=\left[H_{0}(\mathbf{r},t)+D({\bf r},t)\right]\psi({\bf r},t),
\label{eq:tdse-general}
\end{equation}
with $\psi({\bf r},t)$ the system wavefunction and $D({\bf r},t)$ the interaction operator between the system 
and the external field, in the dipole approximation. In our present numerical implementation we choose a spherical coordinate system for the active electron. We represent the angular variables
in a basis of spherical harmonics and write the wave function as,
\begin{equation}
\psi(\textbf{r},t)=\sum_{l=0}^{\infty}\sum_{m=-l}^{l}\frac{f_{lm}(r,t)}{r}Y_{lm}(\theta,\phi),
\label{eq:wf_lm}
\end{equation}
where the spin-variables of the wavefunctions are ignored.
In an actual calculation we must truncate the spherical harmonics
expansion at some maximum value $L$. In the remaining formulas we abbreviate
the truncated double summation by $\sum_{lm}$.

The time-propagation of the wave function proceeds in spherical coordinates
as follows. Substituting Eq. \eqref{eq:wf_lm} in Eq. \eqref{eq:tdse-general}
and projecting onto the spherical harmonic basis $Y_{lm}(\hat{r})$
we obtain the following coupled differential equations for the radial
channel functions as, 
\begin{equation}
i\frac{\partial}{\partial t}f_{lm}(r,t) =
\hat{h}_{l}(r)f_{lm_{l}}(r,t)
+
\sum_{l^{\prime}m_{l}^{\prime}}
\hat{D}_{lm_{l},l^{\prime}m_{l}^{\prime}}(r,t)f_{l^{\prime}m_{l}^{\prime}}(r,t).
\label{eq:tdse_radial}
\end{equation}
 For the special case of linearly polarized light along the $z$-axis
and in dipole approximation the radial time-evolution operators are
given by,
 \begin{subequations} 
\label{eq:tdse-potentials} 
\begin{align}
\hat{h}_{l}(r) & =-\frac{1}{2}\frac{d^{2}}{d r^{2}}+\frac{l(l+1)}{2r^{2}}+V(r),
\label{eq:tdse-potentials_1}\\
\hat{D}_{lm_{l};l^{\prime}m_{l}^{\prime}}(r,t) & = -i\frac{A(t)}{c}
\delta_{m_{l},m_{l^{\prime}}}  
K_{ll^\prime}(m_l) \hat{t}_{ll^{\prime}}(r),
\label{eq:tdse-potentials_2}\\
K_{ll^\prime}(m_l)& = \delta_{ll\pm1}\sqrt{ \frac{l_{>}^{2}-m_{l}^{2}}{4l_{>}^{2}-1} },
\label{eq:tdse-potentials_3}
\end{align}
 \end{subequations}
with $l_{>}=\max(l,l^{\prime})$ and $\hat{t}_{ll^\prime}$ the radial dipole operator. 
The time-dependent radial dipole operator is given as,
\begin{equation}
\hat{t}_{ll^{\prime}}(r)=\frac{\partial}{\partial r}+(l-l^{\prime})\frac{l_{>}}{r},
\label{eq:dv}
\end{equation}
in the velocity form. The quantity ${\bf A}(t) =- c \int_0^t  dt^\prime {\bf E}(t^\prime)$
represents the field potential in the Coulomb gauge. Within the present context the interaction operator couples atomic
states of equal magnetic quantum number, hence we drop the dependence on $m_{l}$ in the subsequent formulation.

By properly arranging the radial channel functions $f_{l}$
according to their angular momentum label we form the radial vector wavefunction
${\bf F}$. In this case the matrix representation of the TDSE [Eq. (\ref{eq:tdse_radial})] is written as,
 \begin{equation}
\dot{{\bf F}}(t)=-i{\bf H}(t){\bf F}(t),
\label{eq:tdse_fd_matrix}
\end{equation}
where $\dot{{\bf F}}\equiv d{\bf F}(t)/dt$ and 
\begin{equation}
{\bf H}(r,t)=\left[
\begin{array}{ccccc}
\hat{h}_{0} & \hat{D}_{01} & 0 & ... & 0\\
\hat{D}_{10} & \hat{h}_{2} & \hat{D}_{12} & ... & 0\\
0 & \hat{D}_{21} & \hat{h}_{3} & ... & ...\\
... & ... & ... & ... & ...\\
... & ... & ... & \hat{h}_{L-1} & \hat{D}_{L-1,L}\\
0 & 0 & ... & \hat{D}_{L,L-1} & \hat{h}_{L}
\end{array}
\right]. .
\label{eq:h_matrix}
\end{equation}

\section{R-matrix eigenstates in the inner region}
\label{sec:a1}
\subsection{Hamiltonian operator in the inner region}
In the inner region $[0,b]$ the radial wavefunctions $f_{l}(r,t)$
are expanded over the eigenstates of the radial Hamiltonian:
 \begin{equation}
\tilde{h}_{l}=\hat{h}_l + \hat{L}_h \qquad l=0,1,..,L.\label{eq:hl}
\end{equation}
with $\hat{L}_h$ the radial Bloch operator,
\begin{equation}
\hat{L}_{h}=\frac{1}{2}\delta(r-b)\frac{d}{dr}
\label{eq:bloch-h}
\end{equation}
and $\hat{h}_l$ given by Eq. (\ref{eq:hydrogen_hamiltonian}). 
The eigenstates of the R-matrix Hamiltonian operator $\tilde{h}_l$ are uniquely determined if we set 
the boundary conditions needed to be fullfiled at the boundaries $r=0$ and $r=b$. In the
present case on physical considerations we take all solutions to vanish at the origin while at $r=b$ the solutions take non-vanishing values. This choice
makes the radial R-matrix operator Hermitian over the inner region $[0,b]$. 
Therefore for each value of the angular momentum we solve the following eigenvalue problem:
\begin{equation}
\tilde{h}_{l}P_{kl}(r)=\epsilon_{kl}P_{kl}(r),\qquad l=0,1,..,L,
\label{eq:h0}
\end{equation}
where $k$ is an integer labelling the eigenstate. The above eigenvalue differential equation is transformed
to solving a matrix diagonalization problem by employing a B-spline
basis set of size $n_{b}$, order $k_{b}$  \cite{deboor:1978} for the representation
of the solutions $P_{kl}(r)$ in region $I$:
 \begin{equation}
P_{kl}(r)=\sum_{j=2}^{n_{b}}C_{j}^{(kl)}B_{j}^{(k_{b})}(r),\qquad0 \le r\le b.
\label{eq:bsplines}
\end{equation}
In the expansion the first B-spline [$B^{(k_b)}_1(r)$] is excluded
in order to conform to the boundary condition at the origin $P_{kl}(0)=0$.
Note that by definition of the B-splines the amplitude of the eigenstates
on the boundary [$w_{kl}=P_{kl}(b)$] is simply the last coefficient
in the expansion, namely, $w_{kl}=C_{n_{b}}^{(kl)}$. All required
integrals are evaluated, with the Gaussian quadrature rule, to machine
accuracy.

For each partial wave $l=0,1,...,L$ the solutions constitute an orthonormal
basis with $n_b-1$ members,
\[
\sum_{k=1}^{n_b-1}|P_{kl}\rangle\langle P_{kl}|=1,\qquad\langle P_{kl}|P_{k^{\prime}l}\rangle=\delta_{kk^{\prime}},
\]
with real eigenvalues $\epsilon_{kl}$.

\subsection{Dipole operator in the inner region}
While the velocity form of the radial dipole operator includes a first-order derivative term [Eq. (\ref{eq:dv})] 
which taken together with the non-vanishing values of the eigenstates $P_{kl}(r)$ at the boundary $b$ makes 
it non-hermitian.We can make this operator Hermitian by adding the dipole Bloch
operator for the first-order derivative in a similar way as done 
for the field-free Hamiltonian $\hat{h}_l$. Thus if we define the dipole velocity operator in region $I$ as: 
\begin{equation}
\hat{L}_d = \frac{1}{2} \delta(r-b) \cos \theta
 \label{eq:bloch-d}
\end{equation}
we find for the radial velocity operator:
 \begin{equation}
\!\!\!\! \tilde{t}_{kl,k^{\prime}l^{\prime}}= \int_{0}^{b}drP_{kl}(r)\left[ t_{kl,k^\prime l^\prime}-\frac{1}{2}\delta(r-b)\right]P_{k^{\prime}l^{\prime}}(r).
\label{eq:dv_hermitian}
\end{equation}

\section{Taylor propagator}
\label{sec:a3} 
The forward evolution of a time-dependent function ${\bf F}(t)$ from a time $t$ to a time $t+\tau$ by the time-step
$\tau$, can be approximated by the Taylor expansion \cite{smyth:1998}:
 \begin{equation}
F(t+\tau)=\sum_{p=0}^{P}a_{p}\cdot\frac{\partial^{p}}{\partial t^{p}}F(t),
\label{eq:explicit}
\end{equation}
 with $\tau=t_{n+1}-t_{n}$, $n=0,1,....,N$ and $a_{p}=\tau^{p}/p!$.
The above propagation scheme consists of an explicit one-step scheme of order $P$.

When the evolution equation for the $F(t)$ is known as $\dot{F}(t)=-iH(t)F(t)$
the above expression can also be obtained as the $P-$order expansion
of the evolution operator $\exp(-iH(t)\tau)$:
 \begin{eqnarray}
\label{eq:taylor_sum}
F(t+\tau) &=& e^{-i\int_{t}^{t+\tau}dt^{\prime}H(t^{\prime})}F(t) \\
 &\equiv&e^{-iH(t)\tau}F(t)=\sum_{p=0}^{P}\frac{(-i\tau)^{p}}{p!}H^{(p)}F(t). \nonumber
\end{eqnarray}
 The above approximate expressions for the time evolution assumes
that the characteristic time evolution of the Hamiltonian $H(t)$
is much larger than the time step $\tau$. In other words, the Hamiltonian
operator is assumed constant within the interval $[t,t+\tau]$ evaluated
at time $t$. Furthermore, in this summation higher-order time derivatives $(\dot{H}(t), \ddot{H}(t),...)$  of the 
Hamiltonian operator have been dropped, a procedure  very well justified for the electric field strengths used in 
this work.

\bibliographystyle{apsrev} 
\bibliography{/users/lnikolopoulos/doc/mypapers/bib/Thesis}

\end{document}